\documentclass[a4paper]{mem}
\usepackage{natbib}
\usepackage{graphicx}
\usepackage[a4paper]{hyperref}
\begin{document}
   \title{Exploring the Divisions and Overlap between AGB and
Super-AGB Stars and Supernovae.}

   \author{J. J. Eldridge and C. A. Tout\\ }

\institute{University of Cambridge, Institute of Astronomy, The
Observatories, Madingley Road, Cambridge CB3 0HA, England}

\mail {John Eldridge \email{jje.ast.cam.ac.uk}}

\abstract{We discuss the mass ranges over which we find AGB and
super-AGB stars. The most massive super-AGB stars are candidate
progenitors for type~II core-collapse SNe. We discuss the two
supernovae, SN1980K and SN2003gd that provide some restrictions on the
upper mass limit of super-AGB stars.}

\authorrunning{J. J. Eldridge \& C. A. Tout.}  \titlerunning{AGB and
   Super-AGB Stars and Supernovae}  \maketitle

%
\section{Introduction}

The primary factors that affect a stars evolution are initial mass,
initial metallicity, mass-loss, duplicity and extra
mixing above and beyond that from convection described by mixing-length
theory. The interplay between these factors determines the
evolutionary path taken and whether a star gives rise to a
supernova (SN) or produces a white dwarf.  These factors are linked.
A more massive star experiences more mass loss, as does a star with
initially higher metallicity.

We have been studying these factors to determine their effect on the
population of SNe progenitors \citep{et1,eldridge1}. By using
observations of SN types and WR ratios we have been able compare a
number of mass-loss prescriptions and decide on a best fitting
prescription. One interesting aspect of our work is the nature of the
lowest-mass stars that explode as SNe. \citet{IBEN2} suggested
that these would be super-AGB stars which are
similar in structure to AGB stars but have oxygen/neon (ONe) cores
rather than carbon/oxygen (CO) cores.

In this paper we give details of the mass ranges over which we find the different
courses of evolution, AGB, super-AGB and SN.  We also discuss
observational evidence from SNe on super-AGB evolution and the lowest
mass for a SN to occur.  We find that the nature of the lowest mass SN
progenitors may provide a tight constraint on convection in stellar
interiors.

\section{The Mass Ranges of AGB Evolution}

AGB stars result from two courses of evolution. For lower masses the
helium burning shell catches up with the hydrogen burning shell. While
for more massive AGB stars (above $3\,M_{\odot}$),
the convective envelope penetrates down to
the CO core at second dredge up.  Both cases put the hydrogen and helium burning shells in
close proximity, an unstable arrangement that leads to thermal pulses
rather than steady shell burning. Furthermore, because the hydrogen
burning shell is now at a higher temperature near the helium burning
shell, hydrogen burning occurs at a much greater rate and the
luminosity of the star rises by a factor of at least 10.

Most AGB stars have CO cores and do not ignite carbon there.  These go
on to form CO White Dwarfs as the envelope is removed by a stellar
wind during the thermally pulsing phase. If the envelope is not
removed these stars would eventually approach the Chandrasekhar mass ($M_{\rm
Ch}$) and carbon would ignite in the centre
leading to a thermonuclear SNe. The resultant SNe would have
observational characteristics of type~Ia SNe with a type~II
masks.  They are often referred to as a type~$1 \frac{1}{2}$
\citep{type1.5}.

However stars above about $7\,M_{\odot}$ experience carbon burning
\citep{IBEN1}.  The lowest-mass stars in
this range ignite carbon in a shell around the more degenerate central
regions. This shell slowly burns inwards, raising the degeneracy as it
goes.  In more massive stars carbon ignites in the core and then burns
outwards in brief shell flashes rather than in a steady burning
shell. This is due to the mild degeneracy of the CO core when the
burning reaction takes place. Carbon burning leaves an ONe core and
the star becomes a super-AGB star. The carbon burning continues
outwards until it reaches the base of the helium burning shell at
which point it extinguishes. However the hydrogen and helium shells
can still continue to burn outwards until $M_{\rm Ch}$ is reached. At
this point the core collapses but now there is no carbon to ignite and
prevent collapse. Densities can be reached of around $9.6 <\log_{10}
(\rho / {\rm g \, cm^{-3}}) < 9.8$ at which electron capture on to
$^{24}$Mg can occur. This accelerates the core collapse and formation
of a neutron star in a type~II SN. The only difference to other SNe is
that the core collapse is via electron capture rather than iron
photodisintigration.

Our work uses the Cambridge stellar evolution code written by
\citet{E71} and most recently updated by \citet{P95} and
\citet{etopac}. We include convective overshooting as described by
\citet{SPE97}. Because of new nuclear reactions rates and opacity
tables the mass ranges for different behaviour are shifted relative to
previous studies. We list the ranges when we include convective
overshooting ($M_{\rm OS}$) and when we remove it ($M_{\rm noOS}$). The
behaviour is as follows.

\begin{itemize}
\item $M_{\rm noOS} \le 7\,M_{\odot}$, $M_{\rm OS} \le 5\,M_{\odot}$ --
stars undergo second dredge-up and thermal pulses with a central CO
core.  They lose their envelopes and
leave CO white dwarfs before before their cores reach $M_{\rm Ch}$.
\item $M_{\rm noOS} \approx 8\,M_{\odot}$, $6\,M_{\odot} \le M_{\rm
OS} \le 7\,M_{\odot}$, during or after second dredge-up carbon
ignites in a shell because the core is degenerate and has a
temperature inversion caused by neutrino losses. The star then
undergoes thermal pulses with an ONe core.  Because the core mass after
dredge-up is far from $M_{\rm Ch}$ we assume these stars lose their
envelopes and form ONe white dwarfs.  These are super-AGB stars.
\item $M_{\rm noOS} \approx 9\,M_{\odot}$, $M_{\rm OS} \approx
7.5\,M_{\odot}$ -- carbon ignites before second dredge-up, in a shell
if the centre is degenerate). Thus at dredge-up there is a growing ONe
core. If this can reach $M_{\rm Ch}$ before the envelope is lost then
the star undergoes a SN. The outcome depends on the nature of the
thermal pulses and the core mass after dredge-up. Only those with cores
close to $1.38\,M_{\odot}$ undergo SNe.
\item $M_{\rm noOS} \approx 10\,M_{\odot}$, $M_{\rm OS} \approx
8\,M_{\odot}$ -- the CO core is greater than $M_{\rm Ch}$ before
dredge-up. However shell carbon burning, enhanced by a thin neon
burning shell in the most massive stars of this type, drives a
convection zone that reduces the size of the CO core to $M_{\rm Ch}$
so dredge-up can occur. This CO material is mixed with the envelope
and increases the CO abundance at the surface during second
dredge-up. After dredge-up the star has an ONe core of $M_{\rm Ch}$
which progresses to a SN by electron capture on to ${\rm Mg}^{24}$. These
are extreme super-AGB stars.
\item $M_{\rm noOS} \ge 11\,M_{\odot}$, $M_{\rm OS} \ge 9\,M_{\odot}$ --
the helium core or CO core masses are too great for dredge-up to
occur. The limiting mass for the helium core is $3\,M_{\odot}$ and for
the CO core $1.5\,M_{\odot}$. Nuclear burning in these stars
progresses until it cannot support the core and a SN explosion ensues.
\end{itemize}

If super-AGB stars do undergo SN then the luminosity of the
lowest-mass SN progenitors depends upon the occurrence of second
dredge-up in the late stages of evolution \citep{Smartt2002}. In the
mass range of interest second dredge-up can increase the final
luminosity from $\log_{10} (L/L_{\odot}) < 4.6$ to $\log_{10} (L/L_{\odot}) >
5.2$. Therefore these stars, if they occur in nature, should provide a
large population of luminous red giant progenitors. The more massive
models increase in luminosity 100 down to~$10\,$yr before the SN. This is
because, before second dredge-up, the stars have more massive cores.
This leads to denser cores after second dredge-up with less time
required to achieve the densities for electron capture. We take this
to occur at a central density of $\log_{10} (\rho / {\rm g \, cm^{-3}}) =
9.8$ \citep{ecapture1,ecapture2}. We
define this as the point when the SN explodes.

Most super-AGB stars probably do not undergo SNe.  They lose their
envelopes and form ONe white dwarfs before core collapse can
occur. Their much higher luminosities enhance mass loss and, in our
models, it can take $100{,}000\,$yr or longer for the burning shells
to advance and for the core to reach the conditions for core collapse
if the core is smaller than $1.38\,M_{\odot}$. To lose their envelopes
during this time a mass-loss rate of only $<10^{-4}\,M_{\odot}\,{\rm
yr^{-1}}$ is required and this is not unreasonable for a luminous red
giant. Also if the helium-burning shell is unstable and thermal pulses
do occur the burning shell's advance is slower and there is even more
time for mass loss.

\begin{figure*}
   \centering
   \resizebox{\hsize}{!}{\rotatebox[]{0}{\includegraphics{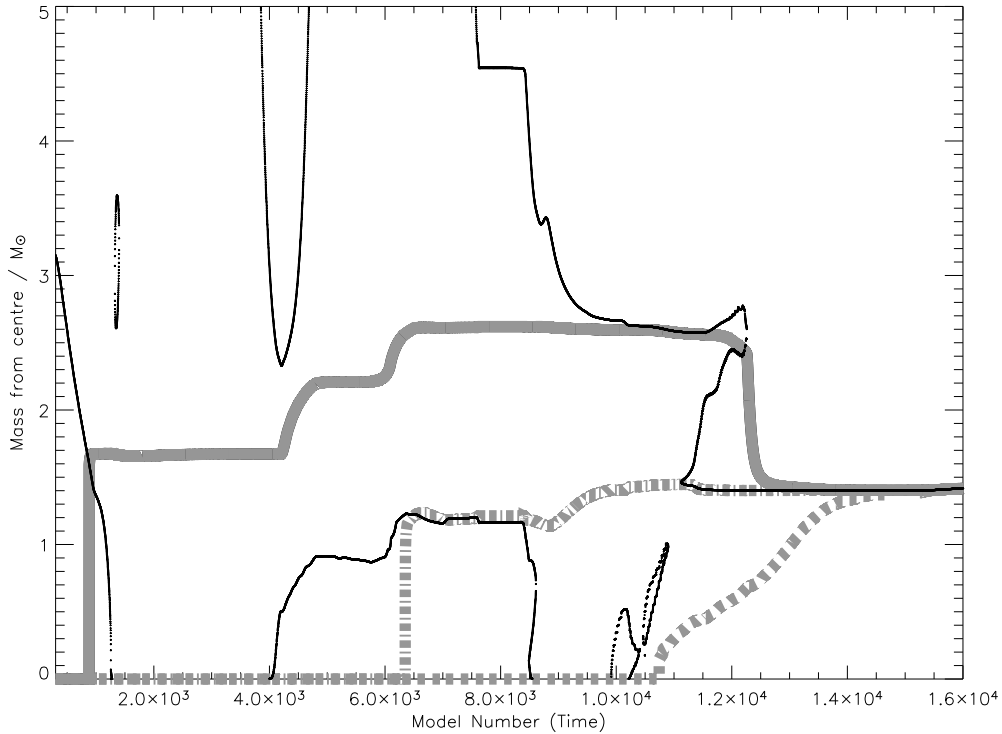}}}
   \caption{Evolution of the structure of a $10\,M_{\odot}$ star.  The
thin black lines show the convective regions while the thick grey lines
represent the burning shells. The solid grey line is the hydrogen-burning shell, the dash-dotted line the helium-burning shell and the dashed line the carbon-burning shell. Notice the formation of the convection zone between the hydrogen and helium burning shell at around model 11,000 that reduces the mass of the CO core before second dredge-up.}
   \label{sagb1}%
   \end{figure*}

However there are a class of stars, extreme super-AGB stars, that have
CO cores greater than $M_{\rm Ch}$ when second dredge-up
occurs. Figure \ref{sagb1} shows an example of such a star at solar
metallicity and without convective overshooting. An intershell convective
region forms and reduces the CO core mass before the convective envelope
penetrates into this region when it mixes helium-burning products to the
surface of the star. Whether these stars exist or some process
prevents second dredge-up is unknown. Only evidence from observations
of SN progenitors will show whether these stars occur in nature.

Eventually the star becomes too massive for second dredge-up to occur
before core collapse either by electron capture or photodisintigration
after the formation of an iron core. However this change over
in behaviour and absence of second dredge-up provides a sharp change in
the luminosity of SN progenitors.

\section{SN Observations}

   \begin{figure*}
   \centering
   \resizebox{\hsize}{!}{\rotatebox[]{-90}{\includegraphics{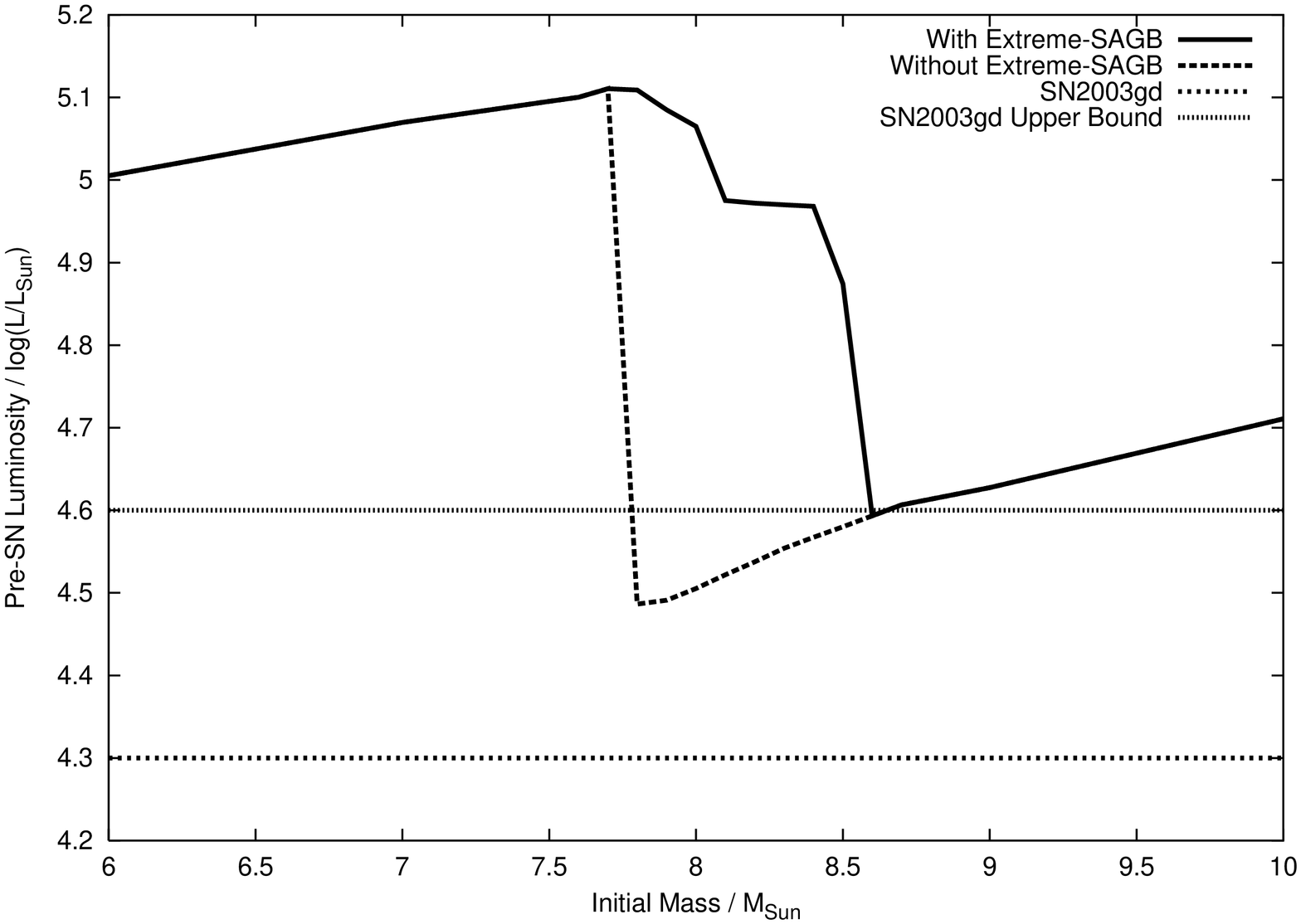}}}
   \caption{The luminosity of SNe progenitors. The solid line includes
the extreme SAGB stars while the dashed line does not. The horizontal
lines are the observed luminosity of the progenitor of 2003gd and its
upper error bar.}
   \label{sagb2}%
   \end{figure*}

There are two SNe that provide evidence for the upper mass for
super-AGB evolution and that super-AGB may undergo SNe. An important
question to ask is what type of SN would a super-AGB stars give rise to? 
Extreme super-AGB stars undergo second dredge-up only just before the SN
and so have little time to lose their envelopes.  They probably give
rise to type~IIP SN. Between these and the stars that have time to
loose their envelopes in a wind there is a range of super-AGB stars
that would lose a large fraction but not all of their envelope and will
give rise to a IIL~SN.

In our calculations we do not see any thermal pulses.  However we do
use low resolution models in this case.  This only affects the stars
that have CO core masses less than $M_{\rm Ch}$.  The super-AGB stars
that we assume explode as SNe have cores that collapse very soon after
second dredge-up.  No further burning needs to take place for the SN
to occur.

First we consider SN1980K, a type~IIL SN, that could have had a
super-AGB progenitor. IIL~SN progenitors have lost most of their
hydrogen envelopes so they cannot sustain a plateau phase in their
light curve. Most progenitors are thought to evolve from massive
single stars that have lost a large fraction of their original
envelopes, either in winds or by a binary interaction.  SN1980K's
progenitor was not observed in pre-SN data but a limit could be placed
on its initial mass of less than $20\,M_{\odot}$ \citep{S03}.  With
the models of \citet{et1} we can rule out a massive single star as
progenitor so it is most likely that the progenitor was in a binary
system.

However it has been suggested by \citet{crazy1} that super-AGB stars
might give rise to IIL~SNe. This agrees with the mass limit from
pre-SN observations and there is another observation that provides
further evidence. \citet{crazy2} deduced the mass-loss history from
radio observations. There was an increase in mass loss $10{,}000\,$yr
before the SN.  This is the same period of time a
$7.5\,M_{\odot}$ model with overshooting experiences between second
dredge-up and SN. The mass loss during this time would reduce the
envelope to the small amount required for a IIL~SN.  So, though there are
more likely alternatives from binary evolution, if we hope to
find super-AGB progenitors, we must look at both IIP and~IIL~SNe.

The other SN to consider is SN2003gd. It is the only red giant
progenitor observed with a pre-SN luminosity of $\log_{10}
(L/L_{\odot}) = 4.3 \pm 0.3$. If we match this against our models,
solid curve in
Figure \ref{sagb2} we can see that they only just agree with the
upper error bar.  We know that the progenitor did not go through second
dredge-up because the pre-SN observations were taken less
than one year before the SN and our models in this range undergo
second dredge-up 10 to $100\,$yr before the SN.  The dashed curve
in Figure \ref{sagb2} is what we would expect the
luminosity of SN progenitors to be if second dredge-up were not to
occur once the CO core is greater than $M_{\rm Ch}$ so that no extreme
super-AGB stars would exist.  These agree slightly better with
SN2003gd so we might need to prevent second dredge-up in these
stars.

A simple way to prevent dredge up would be to restrict convection at
the base of the envelope so that dredge up does not occur and the core
collapses before this. Other possible variables that are likely to
have an effect are the $^{12}$C$+\alpha \rightarrow$ $^{16}$O reaction
rate and neutrino cooling rates. Convection itself is very uncertain and
models are simple.  In future we can refine our theory and use
observations of SN progenitors similar to that of SN2003gd to provide
a limit on the nature of convection in stellar interiors.

\section{Future Work}
Our future plans include exploring the effect of convection on
lowest-mass SN progenitors and looking for low-luminosity progenitors
in models of binary stars.

\begin{acknowledgements}
JJE would like to thank PPARC for his studentship and Fitzwilliam
College for his scholarship.  CAT thanks Churchill College for his fellowship.
\end{acknowledgements}

\bibliographystyle{aa}

\end{document}